# Ab-initio Mapping of Projected Local Density of States in Arbitrary Nanostructures: Application to Photonic Crystal Slabs and Cavities

Gengyan Chen, Yi-Cong Yu, Xiao-Lu Zhuo, Yong-Gang Huang[†],
Haoxiang Jiang, Jing-Feng Liu[‡], Chong-Jun Jin, and Xue-Hua Wang*

State Key Laboratory of Optoelectronic Materials and Technologies, School
of Physics and Engineering, Sun Yat-sen University, Guangzhou 510275, China

## Abstract

Based upon projected local density of states (PLDOS) for photons, we develop a local coupling theory to simultaneously treat the weak and strong interaction between a quantum emitter and photons in arbitrary nanostructures. The PLDOS is mapped by an extremely flexible and efficient method. The recent experiment observation for the photonic crystal slabs is very well interpreted by our *ab-initio* PLDOS. More importantly, a bridge linking the PLDOS and cavity quantum electrodynamics is for the first time established to settle quality factor, g factor and vacuum Rabi splitting. Our work greatly enriches the knowledge about the interaction between light and matter in nanostructures.

Controlling interaction between a quantum emitter and photons at the nanoscale has been central subject of nano-optics with intense activities. Some prominent examples include modification of spontaneous emission rates[1, 2, 3], vacuum Rabi splitting[4, 5], lasing under strong coupling[6], single-photon source[7] and Anderson localization[8]. The interaction may be characterized by the local coupling strength (LCS)[9, 10] proportional to projected local density of states (PLDOS)[11, 12]. Hence tailoring the PLDOS plays a key role in controlling interaction at the nanoscale.

Due to the pivotal role of the PLDOS, the probe of the PLDOS via spontaneous emission rate in various kinds of nanostructures has recently received special attention, such as diamond-structured photonic crystal (PC)[3], random photonic media[13], disordered metal film[14], metal nanowires[15] and PC slab[16]. However, the quantitative theory explanations for the results have been still lacking due to the challenge of simulating the PLDOS in arbitrary nanostructures. Furthermore, this probe approach is valid only for the case of weak coupling between a quantum emitter and photons. In this case, the spontaneous emission rate, i.e. the inverse of spontaneous emission lifetime, is just equal to the LCS at transition frequency between two levels of the quantum emitter[9], and then the PLDOS can be obtained by the proportional relation between the spontaneous emission rate and it.

On the other hand, the solid-state cavity quantum electrodynamics (CQED) systems with strong coupling interaction between a quantum emitter and cavity mode have been a research focus, because they not only provide test beds for fundamental quantum physics but also have important applications in quantum information processing[4, 5, 17]. In strong coupling regime, there is reversible exchange of a single photon between the quantum emitter and cavity mode. The spontaneous emission rate cannot describe this dynamic process and the above-mentioned probe approach of the PLDOS is hence invalid in the strong coupling systems. Certainly, it is a vital demand to establish a linking bridge between the PLDOS and the CQED for both the probe of the PLDOS and manipulation of quantum natures of the solid-state CQED in the strong coupling regime. Up to now, the linking bridge is still an open question.

Motivated by the above-mentioned vital challenge and demand, we develop an extremely flexible and efficient method to ab-initio map out the PLDOS in arbitrary nanostructures, and for the first time establish the linking bridge between the PLDOS and the CQED, which enables the local coupling strength theory simultaneously treat the spontaneous emission in the weak coupling region and the CQED in the strong coupling region. Firstly, the validity of the method is tested in single silver nanosphere. Then, the ab-initio PLDOS of the PC slab samples recently investigated by Wang et al. [16] are mapped out, and they are in good agreement with the probed PLDOS. It is found that the spontaneous emission lifetime of quantum emitter in PC slab is strongly dependent on the orientation of transition dipole moment, and the PC slab has no gap inhibition effect for transition dipole moment being normal to the slab. More importantly, we establish a linking bridge between the PLDOS and the CQED to determine the quality factor, g factor and vacuum Rabi splitting which characterize the CQED. The measured results in the pioneering experiment about the solid-state strong-coupling system between a quantum dot and PC L3 cavity[4] are for the first time reproduced from the ab-initio data of the PLDOS.

The interaction between photons and two-level quantum emitter in a nanostructure is characterized by local coupling strength as[9, 10]:

$$\Gamma(\mathbf{r}_0, \omega) = 2\pi \sum_\lambda |g_\lambda(\mathbf{r}_0)|^2 \delta(\omega - \omega_\lambda) \tag{1}$$

where $\mathbf{r}_0$ is the location of quantum emitter;

$$g_\lambda(\mathbf{r}_0) = i\omega_0 (2\varepsilon_0 \hbar \omega_\lambda)^{-1/2} \mathbf{d} \cdot \mathbf{E}_\lambda(\mathbf{r}_0) \tag{2}$$

is the coupling coefficient; $\omega_0$ is the transition frequency of the bare quantum emitter from exited state to ground state; $\omega_\lambda$ and $\mathbf{E}_\lambda(\mathbf{r})$ are the frequency and electric field of the $\lambda$-th eigenmode in the nanostructure; $\mathbf{d} = d\hat{\mathbf{d}}$ is transition dipole moment between two levels.

The projected local density of states[11, 12] is defined as:

$$\rho(\mathbf{r}_0, \omega, \hat{\mathbf{d}}) = \sum_\lambda |\hat{\mathbf{d}} \cdot \mathbf{E}_\lambda(\mathbf{r}_0)|^2 \delta(\omega - \omega_\lambda), \tag{3}$$

It is straightforward to obtain the LCS as:

$$\Gamma(\mathbf{r}_0,\omega) = \Gamma_0 \frac{\omega}{\omega_0} M(\mathbf{r}_0,\omega,\hat{\mathbf{d}}) \tag{4}$$

Here $M(\mathbf{r}_0,\omega,\hat{\mathbf{d}}) = \rho(\mathbf{r}_0,\omega,\hat{\mathbf{d}})/\rho_0(\mathbf{r}_0,\omega)$ is multiplication factor of PLDOS, i.e. the normalized PLDOS to the density of states (LDOS) $\rho_0(\mathbf{r}_0,\omega) = \omega^2/3\pi^2 c^3$ in vacuum. $\Gamma_0 = \omega_0^3 d^2/3\pi\hbar\varepsilon_0 c^3$ is the spontaneous emission rate of quantum emitter in vacuum.

Various methods, such as Green function method[18], Brillouin zone method[19] and finite difference time domain method (FDTD)[20, 21], have been proposed to simulate the PLDOS for exploring the enhancement and inhibition effects of spontaneous emission in PCs. But the fast and efficient simulation of the PLDOS in arbitrary nanostructures has been still a challenge[21]. The following method and technique are developed to overcome the challenge.

The PLDOS can be expressed by dyadic Green's function as[11, 22]

$$\rho(\mathbf{r}_0,\omega,\hat{\mathbf{d}}) = \frac{2\omega}{\pi c^2} \mathrm{Im}\{\hat{\mathbf{d}} \cdot \vec{\mathbf{G}}(\mathbf{r}_0,\mathbf{r}_0,\omega) \cdot \hat{\mathbf{d}}\}. \tag{5}$$

From Maxwell equations, it can be proved that the electric field induced by an oscillating point-dipole $\mathbf{d} = de^{-i\omega t}\hat{\mathbf{d}}$ located at $\mathbf{r}_0$ is[23]

$$\mathbf{E}_d(\mathbf{r},\omega) = \mu_0 \omega^2 \vec{\mathbf{G}}(\mathbf{r},\mathbf{r}_0,\omega) \cdot d\hat{\mathbf{d}} \tag{6}$$

This implies that the PLDOS can be obtained by the electric field of an oscillating point-dipole at its location as:

$$\rho(\mathbf{r}_0,\omega,\hat{\mathbf{d}}) = \frac{2\varepsilon_0}{\pi\omega} \mathrm{Im}\left\{\frac{\hat{\mathbf{d}} \cdot \mathbf{E}_d(\mathbf{r}_0,\omega)}{d}\right\}, \tag{7}$$

which can greatly simplify the calculation of the PLDOS because the electric field of an oscillating point-dipole can be flexibly and efficiently simulated by various numerical methods, such as multiple scattering method, finite element method and FDTD method. More importantly, only the electric field at the location of the point-dipole needs to be stored and processed, which can greatly save computer time and memory. It is noted that Eq. (7) can easily reduce to results for 1D and 2D cases in Ref. [24].

Usually in FDTD method, a Gaussian pulse as the point-dipole source is added [23, 24] to simulate the time evolution of the electric field $\mathbf{E}_d(\mathbf{r}_0, t)$ induced by the point-dipole, then $\mathbf{E}_d(\mathbf{r}_0, \omega)$ can be obtained by Fourier transformation of $\mathbf{E}_d(\mathbf{r}_0, t)$. But this is very time-consuming because Fourier transformation requires very long time data to obtain convergent results. In order to greatly accelerate the calculation, we adopt the Pade approximation with Baker's algorithm[25] instead of Fourier transformation to more efficiently obtain the complex amplitude of electric field $\mathbf{E}_d(\mathbf{r}_0, \omega)$. Especially, the Pade approach is extremely efficient for calculating the PLDOS in nanostructures with highly localized field distribution, such as nano-cavities. Our test for the PC L3 cavity shows the Pade approach can greatly save computation time by about 200 times than the Fourier transformation.

**Validation of the method.** Firstly, we verify the validation of our method by calculating the PLDOS in the vacuum with a single silver nanosphere with radius 20nm. The distance between the piont-dipole and the centre of the nanosphere is 25nm. The orientation of dipole is along radial direction from the center of the nanosphere to the dipole. The frequency-dependent dielectric permittivity of silver is obtained by interpolating the experimental data[26]. Fig. 1 shows the multiplication factor of PLDOS calculated by our numerical method agrees very well with that exactly obtained by Mie scattering theory[27], which validates our method.

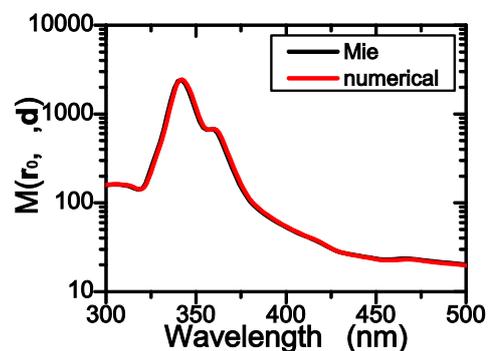

Fig. 1: Multiplication factor of PLDOS in the vacuum with a single silver nanosphere, calculated by Mie scattering theory and our numerical method, respectively.

**The PLDOS in PC slab.** The spontaneous emission lifetime of a quantum emitter in arbitrary nanostructures in weak coupling regime is related by [9]:

$$\tau(\mathbf{r}_0) = \frac{1}{\Gamma(\mathbf{r}_0, \omega_0)} = \frac{\tau_0}{M(\mathbf{r}_0, \omega_0, \hat{\mathbf{d}})} \tag{8}$$

where $\tau_0$ denotes the spontaneous emission lifetime of quantum emitter in vacuum. Apparently, the multiplication factor $M(\mathbf{r}_0, \omega_0, \hat{\mathbf{d}})$ of the PLDOS at $\omega = \omega_0$ in weak coupling regime is reduced to Purcell factor[1].

Recently, many experiments are reported in probing the PLDOS via spontaneous emission lifetime in various kinds of nanostructures[3, 13-16] according to Eq. (8). But the quantitative theory explanations are still lacking due to the difficulty in theoretical mapping of the PLDOS. We now apply our method to map out the *ab-initio* PLDOS in the experimental PC slab samples investigated in Ref. [16], as shown in Fig. 2(a).

Wang *et al.* [16] uses single self-assembled InGaAs quantum dots as internal probes to obtain the PLDOS in GaAs (refractive index n=3.5) PC slabs with circular air holes in triangular lattice. The lattice constant *a* ranges from 200 to 385 nm in steps of 5 nm, while the air hole radius *r* varies along with lattice constant as r=0.3a, and the slab thickness (d) is fixed at d=154nm. A layer of quantum dots are embedded in the slab center and are excited with a pulsed diode laser at 781 nm, and it selects only quantum dots that emit within a narrow spectral range of 970±5 nm. The experiment is very ingenious. They also tried to interpret their experimental results based upon the scaling invariant law from the data of a PC slab sample.

However, according to the scaling invariant law of Maxwell equations[28], only if the PC slabs with different lattice constants have the identical lattice type, refractive index, normalized air hole radius (r/a) and normalized slab thickness (d/a), they have the identical photonic band diagram and the multiplication factor $M(\mathbf{r}_0, \omega, \hat{\mathbf{d}})$ with respect to normalized frequency (a/λ). For the PC slab samples in Ref. [16], the normalized slab thickness d/a decreases as the lattice constant increases, since the slab thickness is fixed at d=154nm. As a result, the photonic band gap should shift to the high normalized frequency[29], rather than keep unchanged. Therefore, the experimental observations cannot be explained from the calculated results of an experimental sample due to the breaking of the scaling invariant law for different samples, as demonstrated below. It is necessary to perform calculations for all of the experimental samples.

Fig. 2(b), (c) and (d) show the multiplication factor $M(\mathbf{r}_0, \omega, \hat{\mathbf{d}})$ of the PLDOS for four different positions on the central plane of the PC slabs with three different lattice constants. It can be observed that the multiplication factor $M(\mathbf{r}_0, \omega, \hat{\mathbf{d}})$ changes by almost two orders of magnitude for four different positions $\mathbf{r}_0$, which means the PLDOS in PC slab and then the spontaneous emission lifetime of quantum dot $\tau(\mathbf{r}_0)$ are strongly dependent on the position[9]. For each slab, there is a deep concave within the same region of the normalized frequency for four different positions, which just corresponds to the photonic band gap of each PC slab where the PLDOS is strongly suppressed. For three different PC slabs, the normalized frequency (the vertical magenta dash lines) of the quantum dot locates below, inside and above the individual band gap, respectively. This indicates that the enhancement or inhibition of the spontaneous emission from a quantum dot can be control by adjusting the lattice constant.

It is worth pointing out that the widths and positions of the photonic band gaps for three slabs with different lattice constants are different. The photonic band gap shifts to the high normalized frequency with increasing lattice constant. No scaling invariant law is observed, as is in accord with the previous analysis.

In order to understand the experimental results in Ref.[16], we calculate the multiplication factors $M(\mathbf{r}_0, \omega_0, \hat{\mathbf{d}})$ of the PLDOS at transition wavelength of 970nm for all of the PC slab samples with the lattice constant increasing from a=200nm to a=385nm by a step of 10nm. The results of the PLDOS in unit of $4/3a^2c$ are shown in Fig. 2(e) for comparing with probed PLDOS in FIG. 2 of Ref.[16]. In order to completely reflect the distribution of the PLDOS in each sample, we firstly search the positions with maximum and minimum values of the electric field. They are respectively $\mathbf{r}_1$=(0.325a, 0, 0) and $\mathbf{r}_4$=(0.025a, 0.4a, 0). We then choose four random positions, corresponding to the magenta dots in Fig. 2(e). As expected, the PLDOS at the random position falls within the PLDOS values at positions $\mathbf{r}_1$ and $\mathbf{r}_4$.

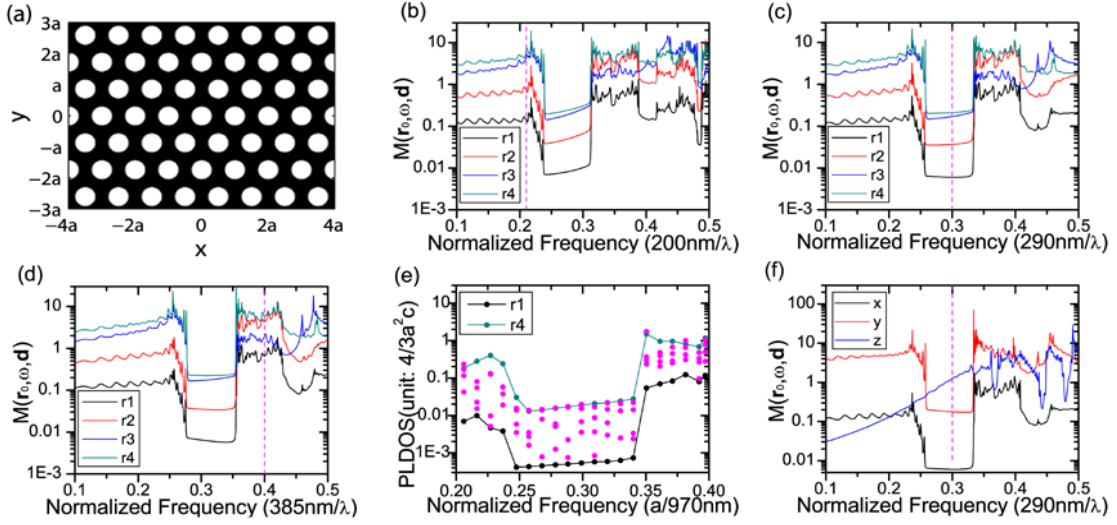

Fig. 2: (a) Cross-section structure sketch on central plane (z=0 plane) of PC slabs, Gray region is dielectric slab and white regions are air holes. (b)-(d) the multiplication factor $M(\mathbf{r}_0, \omega, \hat{\mathbf{d}})$ of the PLDOS in three PC slabs with different lattice constant: (b) for a=200nm, (c) for a=290nm and (d) for a=385nm. The position $\mathbf{r}_0$ is r1=(0.325a, 0, 0), r2=(0.475a, 0, 0), r3=(0.525a, 0.3a, 0) and r4=(0.025a, 0.4a, 0), respectively. The orientation of $\hat{\mathbf{d}}$ is along x direction. The vertical magenta dash lines denote transition wavelength (970nm) of quantum dot. (e) The PLDOS at transition wavelength in PC slabs with different lattice constants. In each PC slab, r1=(0.325a, 0, 0) and r4=(0.025a, 0.4a, 0) correspond to two positions with the minimum and maximum electric fields, respectively. The magenta dots denote four random locations. $\hat{\mathbf{d}}$ is along x direction. (f) $M(\mathbf{r}_0, \omega, \hat{\mathbf{d}})$ in the PC slab with a=290nm. $\mathbf{r}_0$ is (0.325a, 0, 0), and $\hat{\mathbf{d}}$ is along x, y and z direction, respectively.

From Fig. 2(e), we also observe a deep concave in a wide range of the normalized frequency. The width of the deep concave is larger than that of each slab sample. Obviously, the results in Fig. 2(e) reflect the total effect of all slab samples, and are in good agreement with the experimental results in FIG. 2 of Ref. [16] with the transition dipole moment along X direction.

We further investigate the orientation-dependent character of the PLDOS and spontaneous emission lifetime in PC slab. We choose the PC slab with a=290nm, where the transition wavelength is inside the photonic band gap, and calculate the multiplication factor $M(\mathbf{r}_0, \omega, \hat{\mathbf{d}})$ with $\hat{\mathbf{d}}$ along x, y and z direction, respectively. The results are shown in Fig. 2(f). The $M(\mathbf{r}_0, \omega, \hat{\mathbf{d}})$ show that there are band gaps for x and y orientation, while no band gap exists for z orientation, which just reveals the fact that the hole slab favors to form the band gap of the TE-like mode[28]. The multiplication factors of the PLDOS at transition frequency $M(\mathbf{r}_0, \omega_0, \hat{\mathbf{d}})$ are 0.006, 0.177 and 1.221 for x, y and z orientation, respectively. This indicates that the spontaneous emission lifetime of quantum dot in PC slab is strongly dependent on orientation due to pseudo photonic band gap effect.

**Linking bridge between the PLDOS and the CQED.** Unlike the irreversible decay of quantum emitter in the weak coupling regime, in solid-state CQED systems with strong coupling interaction between a quantum emitter and cavity mode, there is reversible exchange of a single photon between the quantum emitter and cavity mode. The spontaneous emission rate can no longer describe this dynamic process and the above-mentioned probe approach of the PLDOS is invalid. It is significant to establish a linking bridge between the PLDOS and the CQED.

As well known, the LCS between a quantum emitter and an ideal single-mode cavity without loss can be expressed as:

$$\Gamma(\mathbf{r}_0, \omega) = 2\pi |g_c(\mathbf{r}_0)|^2 \delta(\omega - \omega_c) \tag{9}$$

where $\omega_c$ is frequency of cavity mode; $|g_c(\mathbf{r}_0)|$ is g factor that characterizes the coupling strength between quantum emitter and single-mode cavity.

For the realistic single-mode cavity with loss, the LCS may be expressed as Lorentz function:

$$\Gamma(\mathbf{r}_0, \omega) = 2|g_c(\mathbf{r}_0)|^2 \frac{\kappa/2}{(\omega - \omega_c)^2 + (\kappa/2)^2} \tag{10}$$

where $\kappa = \frac{\omega_c}{Q}$ is decay rate of cavity, Q is quality factor of cavity. Apparently, when the decay rate is infinitely small, Eq. (10) reduce to Eq. (9).

For a cavity with high quality factor and in resonance with the quantum emitter, from Eq. (4) and (10), we can also express multiplication factor of the PLDOS in cavity as Lorentz function:

$$M(\mathbf{r}_0, \omega, \hat{\mathbf{d}}) = \frac{2|g_c(\mathbf{r}_0)|^2}{\Gamma_0} \frac{\kappa/2}{(\omega-\omega_c)^2 + (\kappa/2)^2} \quad (11)$$

Simply by fitting the *ab-initio* data of $M(\mathbf{r}_0, \omega, \hat{\mathbf{d}})$ with Lorentz function, we can obtain the mode frequency, quality factor of the cavity. g factor of the cavity can be obtained by the peak value of $M(\mathbf{r}_0, \omega, \hat{\mathbf{d}})$ as follows:

$$M(\mathbf{r}_0, \omega_c, \hat{\mathbf{d}}) = \frac{4|g_c(\mathbf{r}_0)|^2}{\Gamma_0 \kappa} = \frac{8}{N_0} \quad (12)$$

Here, $N_0$ is critical atom number and is an important parameter characterizing the CQED[30]. We can hence calculate g factor simply by:

$$|g_c(\mathbf{r}_0)| = \frac{1}{2}\sqrt{\Gamma_0 \kappa M(\mathbf{r}_0, \omega_c, \hat{\mathbf{d}})} \quad (13)$$

From dressed-atom state[10], we can further derive vacuum Rabi splitting as:

$$\Omega = 2\sqrt{|g_c(\mathbf{r}_0)|^2 - (\frac{\kappa}{2})^2} \quad (14)$$

So far, we have established a linking bridge between PLDOS and important parameters characterizing CQED, including quality factor, g factor, vacuum Rabi splitting and critical atom number. Since the calculation of the PLDOS in cavity is extremely efficient by Pade approach, this simple linking bridge enables us investigate solid-state CQED efficiently.

As a demonstration, we investigate a quantum dot in PC L3 cavity following the design of the pioneering experimental solid-state strong-coupling system[4], as shown in Fig. 3(a). The structure is composed of GaAs (n=3.4) with a triangular lattice of air holes with lattice constant a=300nm. The slab thickness is 0.9a and the air hole radius is 0.27a. This PC L3 cavity is made by missing three air holes in a line and displacing two air holes at both edges of the cavity by 0.2a. The quantum dot and PC L3 cavity are tuned to exact resonance. The spontaneous emission lifetime of quantum dot in GaAs slab without PC pattern is 1.82ns.

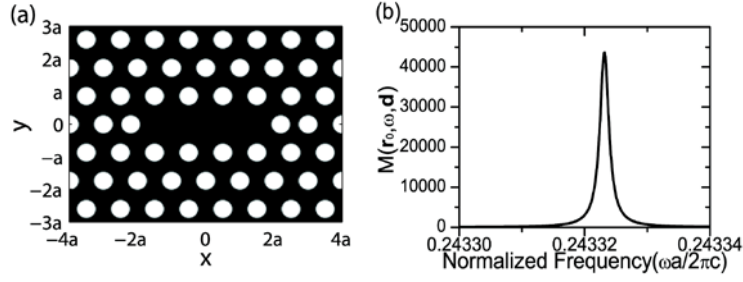

Fig. 3: Quantum dot in PC L3 cavity. (a) Cross-section on central plane (z=0 plane) of PC L3 cavity. Gray region is dielectric slab and white regions are air holes. (b) $M(\mathbf{r}_0, \omega, \hat{\mathbf{d}})$ in PC L3 cavity. $\mathbf{r}_0 = (0,0,0)$ is the center of cavity and $\hat{\mathbf{d}}$ is along y direction.

The $M(\mathbf{r}_0, \omega, \hat{\mathbf{d}})$ in Fig. 3(b) calculated by Pade approach can be very well fitted by Lorentz function of Eq. (11). From this we determine all the characteristic parameters of the CQED system: the normalized frequency of cavity mode is 0.2433232; the quality factor is Q=140398; g factor is g=22.1GHz; the vacuum Rabi splitting is 44.1GHz. The obtained mode frequency, g factor and vacuum Rabi splitting are all in good agreement with experimentally observed values[4], except for the calculated quality factor that is about 8 times larger than experimental value. In order to under the difference in the quality factor, we have recalculated the quality factor of some PC L3 cavities in various references[31], and found excellent agreement with those calculated by other numerical methods. Therefore, the disagreement between theoretical and experimental value of quality factor may be attributed to the fabrication imperfection of PC L3 cavity[32]. The further investigation about the effect of the fabrication imperfection on CQED will be presented elsewhere.

In summary, the local coupling theory based upon the PLDOS has been constructed to simultaneously treat the spontaneous emission in the weak coupling region and the CQED in the strong coupling region. An extremely flexible and efficient method is developed to map out the PLDOS in arbitrary nanostructures. Based upon the *ab-initio* PLDOS, the recent experimental results about the PC slab are very well interpreted. It is also found that the orientation of transition dipole moment have a profound influence on the spontaneous emission of a quantum dot in the PC slabs, and even the PC slab has not any gap inhibition effect when the transition dipole moment being normal to the slab. For the first time, we have established a linking bridge between the PLDOS and the CQED to determine the quality factor, g factor and vacuum Rabi splitting. The measured results in the pioneering experiment about the solid-state strong-coupling system between a quantum dot and PC L3 cavity are for the first time reproduced from the ab-initio data of the PLDOS. Our work greatly enriches the knowledge about the interaction between light and matter in nanostructures, and can provide a guidance to tailoring the interaction between light and matters at the nanoscale.

## Acknowledgment

This work was financially supported by the National Basic Research Program of China (2010CB923200), the National Natural Science Foundation of China (Grant U0934002), and the Ministry of Education of China (Grant V200801).

______________________________________________________________________

\* Corresponding author: wangxueh@mail.sysu.edu.cn

[†] Present address: College of Physics Science and Information Engineering, Jishou University, Jishou 416000, China

[‡] Present address: College of Science, South China Agriculture University, Guangzhou 510642, China

______________________________________________________________________